\documentclass[journal]{IEEEtran}

\usepackage{times}
\usepackage{epsfig}
\usepackage{graphicx}
\usepackage{amsmath}
\usepackage{amssymb}
\usepackage[ruled,vlined]{algorithm2e}
\usepackage{algpseudocode}
\usepackage{caption}
\usepackage{subcaption}
\usepackage{makecell}
\usepackage{booktabs}
\usepackage{tabularx}
\usepackage{xcolor}
\usepackage{bm}
\usepackage{cite}
\usepackage{textcomp}
\usepackage{mathrsfs}
\usepackage{graphicx}
\graphicspath{ {./images/} }

\begin{document}
%
\title{Pilot design, channel estimation, and target detection for integrated sensing and communication with OTFS}
%
%
%

\author{Dazhuo Wang, Yonghong Zeng, Yuhong Wang, Francois Chin, Yugang Ma, and Sumei Sun.\\
	Institute for Infocomm Research (I$^2$R), Agency for Science, Technology and Research (A*STAR), Singapore
\thanks{This research is supported by the National Research Foundation, Singapore, and Infocomm Media Development Authority under its Future Communications Research and Development Program under Grant FCP-NUS-RG-2022-018}

}


\maketitle

\begin{abstract}
Recent studies shows that the orthogonal time frequency space (OTFS) waveform is a promising candidate for future communication. To meet users' potential demand for Integrated Sensing and Communication (ISAC) applications in 6G, the usage of OTFS for both radar sensing and wireless communication needs to be explored. In this paper, we propose a Fast Algorithm OTFS radar (FAOR) that can perform radar sensing in low complexity to detect the range and speed of the targets. It computes the 2D cyclic correlation of transmitted signal with the reordered delay Doppler (DD) domain received signals, and then generates the 2D range-Doppler map. It can be applied not only to monostatic radar but also to bistatic radar with a much lower computational complexity compared to state-of-the-art radar sensing technology. With the detected time delays and Doppler frequencies of the targets after the radar sensing, we propose a pilot-aided channel estimation method. The multifunction pilot symbol can serve the purpose of both bistatic radar sensing and channel estimation without any guard symbol added, while reducing the peak-to-average power ratio (PAPR) considerably compared to the conventional pilot design. The simulation results show that the proposed scheme outperforms the compared algorithms and gives decent performance in both radar sensing and channel estimation.

\end{abstract}


\section{Introduction}

In modern society, people require stable communication anytime and anywhere to meet their increasing demands for work, entertainment, etc. In high mobility scenarios, such as moving vehicles, planes, or satellites, the existing orthogonal frequency division multiplexing (OFDM) modulation technique suffers from the impact of Doppler spread, leading to performance degradation due to the high Doppler shift. As a potential alternative, the orthogonal time frequency space (OTFS) has recently received increasing attention. 

Research \cite{hadani2017orthogonal,yuan2023new,lin2022orthogonal} shows that OTFS is able to combat the impact of Doppler spread and provides decent performance in high-speed scenarios. It maps and processes the communication information in the delay-Doppler domain. By taking advantage of the Zak transform, it generates a spread of data in time frequency domain. As the transmission symbol is distributed across all time frequency resources, OTFS can perform well on high mobility and multiple path wireless channels \cite{raviteja2018interference}. Given that OTFS may occupy an important position in future communication, it is natural that we also want to explore the possibility that it can be applied for sensing applications. 

The separation of communication and sensing may cause huge spectrum, hardware, and energy waste. In recent years, there has been a trend to merge and synergize sensing and communication to form Integrated Communications and Sensing (ICAS), also called integrated sensing and communication (ISAC). With ISAC, sensing availability and accuracy can be greatly enhanced due to the presence of communication network infrastructures. On the other hand, the feedback of sensing can provide real-time information to improve the performance of the communication networks. Besides, the ISAC can save spectrum, energy \& cost and even enable new applications that need communication and sensing at the same time.

Many research works have studied the pilot design and the channel estimation of OTFS. Reference \cite{hashimoto2021channel} provides new insight into the OTFS input-output relation in the delay Doppler (DD) domain as a 2D shaped circular convolution. It also proposes a low-complexity channel equalization method. In \cite{murali2018otfs}, authors propose a pseudo-noise (PN)-sequence based channel estimation method to estimate the Doppler shift of each path, while the drawback is the high computational load. In \cite{raviteja2019embedded}, the pilot pattern and channel estimation scheme was proposed, and it showed that the OTFS with non-ideal channel estimation outperformed OFDM with ideal channel estimation. The authors used a matrix-form pilot scheme to obtain a close-channel estimation performance in \cite{liang2022pilot}. Authors propose a multiple scattered superposition pilot aided channel estimation scheme \cite{liu2023low}, it is able to improve peak-to-average power ratio (PAPR) and channel estimation performance. The work \cite{zhao2020novel} proposes multiple impulses pilot scheme was proposed for channel estimation and achieved a superior performance with the diversity gain brought by the multiple impulses.

The OTFS based radar sensing is also studied recently to explore the ISAC deployment of OTFS in the future communication networks. Reference \cite{gaudio2020effectiveness} proposes a maximum likelihood (ML) algorithm to estimate range and velocity via the OTFS transmission scheme. The authors proves that the OTFS signal can achieve radar estimation performance bound while maintaining a superior communication performance over OFDM. Reference \cite{raviteja2019orthogonal} proposes a matched-filter algorithm to estimate the range and velocity of targets utilizing the OTFS structure.
Reference \cite{zhang2023radar} proposes a two-dimensional (2D) correlation-based algorithm to estimate the fractional delay and Doppler parameters for radar sensing. 

In this paper, we propose a novel Fast Algorithm OTFS radar (FAOR). It is capable of radar sensing in a fast algorithm manner. It computes the 2D cyclic correlation of the transmitted signal with the reordered delay Doppler (DD) domain received signals based on the embedded pilot symbols, and then generates the 2D range-Doppler map for radar sensing. With the detected time delays and Doppler frequencies of the targets after radar sensing, we propose pilot-aided channel estimation methods. Simulations demonstrate that FAOR can provide decent performance in radar sensing with a lower computation complexity compared to state-of-the-art algorithms. Besides, the proposed pilot design has lower PAPR compared to conventional design, while achieving comparable channel estimation performances.

The rest of the paper is organized as follows: Section II illustrate the structure of OTFS modulation and system model. Section III provides the details of the proposed algorithm FAOR. Section IV is on pilot design and channel estimation. Section V shows simulation results and comparison with existing works. Section VI concludes the paper.

\begin{figure*}[htbp]
	\centering
	\includegraphics[width=15cm,height=3.5cm]{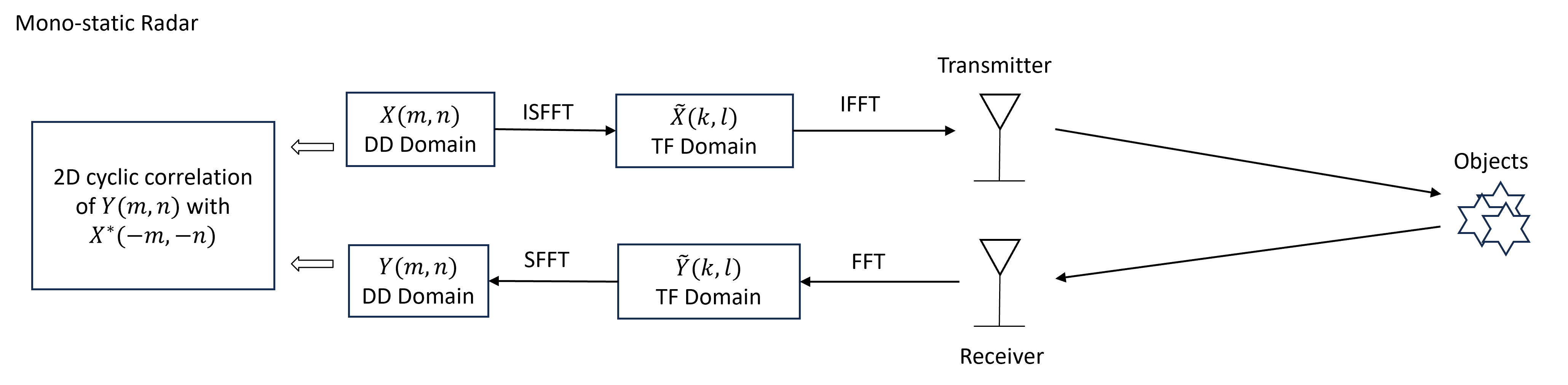}
 \caption{FAOR for Monostatic Radar Sensing}
	\label{fig:mono}
\end{figure*}

\begin{figure*}[htbp]
	\centering
	\includegraphics[width=15cm,height=3.5cm]{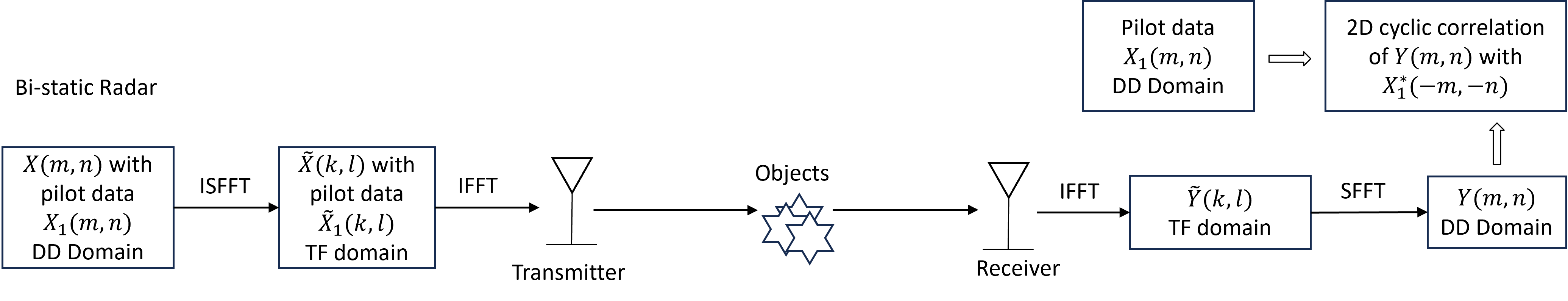}
 \caption{FAOR for Bistatic Radar Sensing}
	\label{fig:bi}
\end{figure*}

\section{OTFS Modulation and System Model}

OTFS modulation was shown to have significant performance advantages over OFDM for doubly selective channels \cite{mohammed2022otfs}.

At the transmitter side, the Delay-Doppler data $X(m,n)$ is converted to Time Frequency domain data $\tilde{X}(k,l)$ through the inverse symplectic fast Fourier transform (ISFFT). Then the signal is transmitted from the transmitter after the Heisenberg Transform. At the receiver side, the receive signal is converted to Time-Frequency Domain $\tilde{Y}(k,l)$. Then it is converted to Delay-Doppler domain data $Y(m,n)$.

The architecture can be simplified by combining the column FFT in the ISFFT and the IFFT in the Heisenberg Transform. The simplified OTFS structure is also called Zak Transform [5]. The Delay Doppler domain data $X(m,n)$ is converted to the Delay Time Domain data $X_{zak}$ through the row IFFT operation. At the receiver side, the received Delay Time Domain data $Y_{zak}$ is converted to Delay Doppler domain data $Y(m,n)$ through the row FFT operation.

In OTFS signal model, the data is placed in delay-Doppler domain. With the number of subcarriers $M$ and number of symbols $N$, the DD domain received signal passed through the channel can be approximated as
\begin{equation}
	\begin{aligned}
		Y(m,n)\approx\sum_{p=1}^{P}h_pe^{j2\pi(\frac{m-l_p}{M})\frac{k_p}{N}}
		a_p(m,n)\\ \cdot X(<m-l_p>_M,<n-k_p>_N)&+\eta(m,n),
	\end{aligned}
\end{equation}
where $a_p$ is defined as follow
\begin{equation}
	a_p(l,k)=
	\begin{cases}
		1, & l_p \le m\le M-1,\\
		\frac{N-1}{N}e^{(-j2\pi \frac{n-k_p}{N})}, & 0\le m\le l_p-1, 
	\end{cases}
\end{equation}
$X(m,n)$ is the DD domain transmitted signal samples, $p$ is the path index of $P$ paths, $l_p$ is the normalized range, $k_p$ is the normalized Doppler frequency, and
\begin{equation}
  l_p=\tau_p/T_s, 0\le l_p\le M-1,
\end{equation}
\begin{equation}
	k_p=Nf_pT=Nf_p/\Delta f,0\le k_p\le N-1,
 \end{equation}
$\Delta f$ is the subcarrier spacing, $T_s$ is the sample period, and $T$ is the duration of one sub-block. Radar sensing is to find the number of reflectors $P$, range $l_p$, and speed $k_p$, for given $Y(m,n)$ and $X(m,n)$. For the purpose of communication, we need to further measure the channel coefficient $h_p$.

\section{Fast Algorithm OTFS radar (FAOR)}

Our proposed algorithm FAOR can be used in both monostatic and bistatic radar sensing. For monostatic radar the entire transmitted signal can be used for sensing, while for bistatic radar only the pilot symbol embedded in the transmitted signal can be utilized.

\subsection{FAOR for Monostatic Radar Sensing}

We firstly introduce the working principle of FAOR in monostatic radar. In monostatic radar, the transmitter and receiver are co-located.

For monostatic radar, the entire transmitted signal can be used for radar sensing. We denote the transmitted symbol as $X(m,n)$ and the received symbol as $Y(m,n)$. Based on the properties of the received signal in the DD domain, we found that the 2D cyclic correlation of the transmitted signal with the reordered received signals can be used to generate the 2D range Doppler map (RDM). The 2D cyclic correlation is defined as
\begin{equation}
    Z(k,l)=\sum_{m=0}^{M-1}\sum_{n=0}^{N-1}Y(m,n)X^*(<m-k>_M,<n-l>_N).
\end{equation}
The 2D cyclic correlation can be computed by 2D FFT. Thus,  FAOR computes the 2D FFT of $X(m,n)$ and denotes the result as $\hat{X}(k,l)$. Then it computes the 2D FFT of $Y(m,n)$ and we denote the result as $\hat{Y}(k,l)$. Then the FAOR computes the $\hat{Z}(k,l)$ as    
\begin{equation}
               \hat{Z}(k,l)=\hat{Y}(k,l)\hat{X}^*(k,l).
\end{equation}
Then we compute the 2D IFFT of the $\hat{Z}(k,l)$, denoted by $Z(k,l)$, which is the 2D RDM. The result $Z(k,l)$ can be used for range and speed detection of targets. Figure \ref{fig:mono} shows the FAOR for monostatic radar.


\subsection{FAOR for Bistatic Radar Sensing}

We then introduce the working principle of bistatic radar sensing. For a bistatic radar, the transmitter and receiver are in a different location. Therefore, the information of transmitted signal is not available at the receiver side. In fact, this is the common situation for the real-word wireless communication networks. To realize the function of ISAC, the information symbol cannot be predefined for radar sensing. Both functions of communication and sensing should be considered. The idea of pilot symbols is applied to address this issue. In the transmitted signal, we add in pilot symbols which is known as prior knowledge at the receiver side. FAOR can just use these pilot symbols for radar sensing.

Similarly as in the case of monostatic radar, FAOR also calculates the 2D cyclic correlation of the signal received from the DD domain $Y(m,n)$ with the signal transmitted from the DD domain $X(m,n)$ and generates the 2D RDM. However, only part of the transmitted signal can be used to preserve the communication function. Instead of using the entire $X(m,n)$, we only use the pilot symbol by setting the pilot grid. These pilot symbols can be treated as prior knowledge on the receiver side. We define $X_1 (m,n)=s_p (m,n)$, if $(m,n)$ is in the pilot grid and $s_p (m,n)$ is the pilot symbol, otherwise, $X_1(m,n)=0$. Then, the FAOR for bistatic sensing is as follows. FAOR computes the 2D FFT of the modified transmitted signal $X_1(m,n)$. The computed result is denoted as $\hat{X}_1(k,l)$. At the received symbols, $\hat{Y}(k,l)$ is computed by taking the 2D FFT of $Y(m,n)$. Then it calculates $\hat{Z}_1(k,l)$ as follows.
\begin{equation}
	\hat{Z}_1(k,l)=\hat{Y}(k,l)\hat{X}_1^*(k,l).
\end{equation}
We then compute the 2D IFFT of the $\hat{Z}_1(k,l)$, denoted by $Z_1(k,l)$, which is the 2D RDM. The result $Z_1(k,l)$ can be used for range and speed detection of targets.  Figure \ref{fig:bi} shows the FAOR for bistatic radar sensing. 


\subsection{Advantages of FAOR}

FAOR is able to do radar sensing in a fast algorithm manner. Compared to the state-of-the-art OTFS radar\cite{zhang2023radar}, it has a lower complexity, as shown in
Table \ref{tab:complexity}.

\begin{table}[htb]
	\begin{center}
		\caption{Complexity Comparison}
		\begin{tabular}{c|c}
			\hline
			\textbf{Method}  &\textbf{Complexity}  \\
			\hline
			FAOR &\makecell[c]{O$(MN{\rm log}_2(MN))$}\\
			\hline
			Algorithm in \cite{zhang2023radar}&\makecell[c]{O$((MN)^2)$}\\
			\hline
			
		\end{tabular}
		\label{tab:complexity}
	\end{center}
\end{table}

\section{Pilot design and channel estimation}

With the range-Doppler map from the sensing result, we can derive the normalized range and Doppler frequency $l_p$ and $k_p$. Based on Equation (1) and the estimated range and Doppler of the targets, we can estimate the channel coefficients. We can reformulate Equation (1) into the following equation
\begin{equation}
    \bold{y}=\bold{X}_{c}\bold{h},
\end{equation}
where $\bold{X}_{c}$ is a combined matrix with all the known elements including Doppler, delay and transmitted pilot symbol with a dimension of $M_{p}N_{p} \times P$, and ${\bf y}$ is a vector of size $M_{p}N_{p} \times 1$ that is formed from the received DD domain signal, where $M_{p}$ and $N_{p}$ is the pilot symbol length and width of the pilot block size, respectively, and $\bold{h}$ is composed of unknown channel coefficients $h_p$ as follows
\begin{equation}
	\begin{aligned}
		\bold{h}=[h_1,\dots,h_p]^T.
	\end{aligned}
\end{equation} 
The elements of matrix $\bold{X}_{c}$ can be computed as follows:
\begin{equation}
	\begin{aligned}
		\bold{X}_{c}(mN+n,p)=e^{j2\pi(\frac{m-l_p}{M})\frac{k_p}{N}}
		a_p(m,n)\\ \cdot X(<m-l_p>_{M},<n-k_p>_{N}).
	\end{aligned}
\end{equation}
Note that only the pilot symbols are used to form the matrix. Then we are able to compute $\bold{h}$ and get the channel estimation as follows
\begin{equation}
	\bold{h}=(\bold{X}_{c}^{\rm H}\bold{X}_{c})^\bold{-1}\bold{X}_{c}^{\rm H}\bold{y}.
\end{equation}

Our designed pilot symbols can be randomly placed anywhere in the transmitted frame, unlike the conventional pilot design. In addition, it does not need any guard symbol.

\section{Simulation}

We have run extensive simulations to test the performance of the proposed algorithms on both radar sensing and channel estimation. 

\subsection{Evaluation on Radar Sensing}

For the test of radar sensing performances, we consider different use cases where both monostatic radar and bistatic radar are applied. We also compare FAOR with other algorithms. The detailed setting is as follows. For the OTFS settings, we set the central frequency as 60 GHz. The number of subcarriers is $M=4096$ and the number of symbols in time is $N=100$. The frame size is around 3.33 ms. The modulation used is 4-QAM constellation. The subcarrier spacing is 30 KHz and the bandwidth is 122.88 MHz.

\begin{figure}[htbp]
	\centering
	\includegraphics[width=8.3cm,height=5cm]{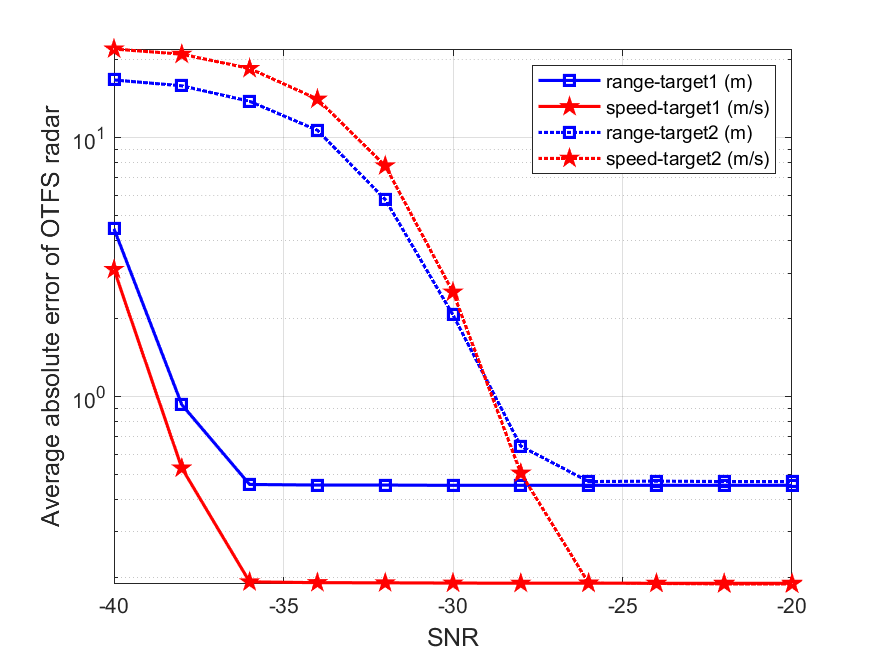}
	\caption{Target Estimation Error of Monostatic Radar Sensing}
	\label{fig:monoee}
\end{figure}

We first test the radar sensing performance of FAOR in monostatic conditions. Figure \ref{fig:monoee} shows the estimation error of the radar sensing on the two detectable targets. The figure shows both the range error and the speed error, recorded from -40 dB to -20 dB. For target 1, the range and speed estimation errors are minimized and stable from -36 dB. For target 2, the range and speed estimation errors are minimized and remain stable from -26 dB.

\begin{figure}[htbp]
	\centering
	\includegraphics[width=8.3cm,height=5cm]{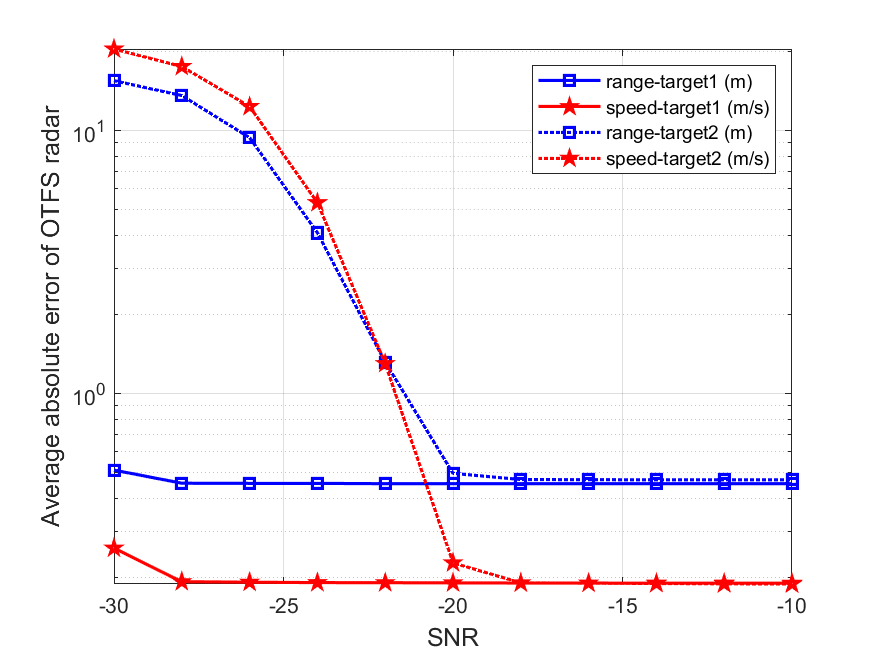}
	\caption{Target Estimation Error of Bistatic Radar Sensing}
	\label{fig:bistaticee18}
\end{figure}

\begin{figure}[htbp]
	\centering
	\includegraphics[width=8.3cm,height=5cm]{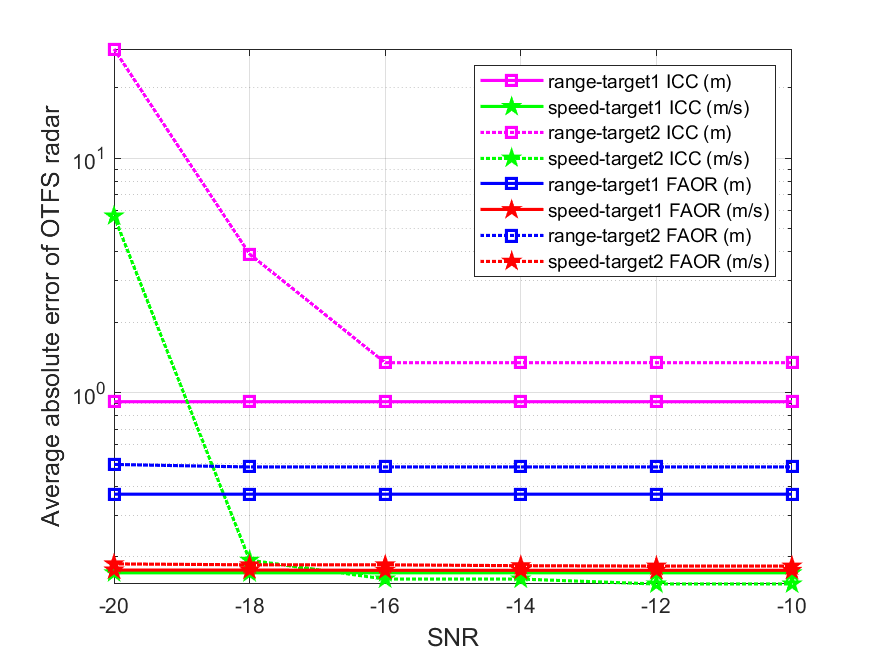}
	\caption{Comparison of Target Estimation Error}
	\label{fig:compareee}
\end{figure}

\begin{figure}[htbp]
	\centering
	\includegraphics[width=8.3cm,height=5cm]{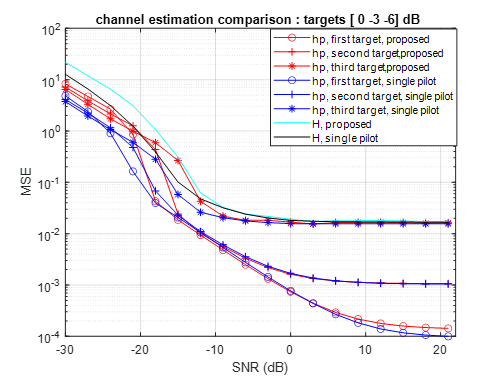}
	\caption{Channel estimation MSE comparison between proposed pilot design and single pilot method in scenario 1}
	\label{fig:s1}
\end{figure}

\begin{figure}[htbp]
	\centering
	\includegraphics[width=8.3cm,height=5cm]{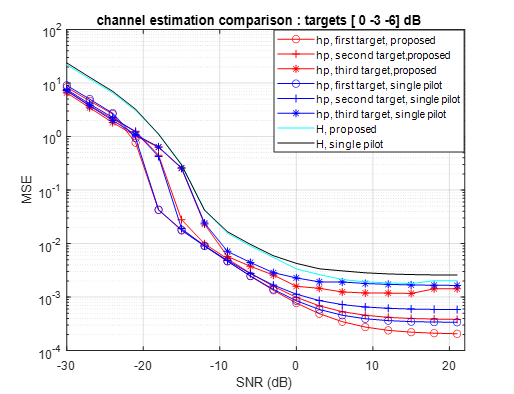}
	\caption{Channel estimation MSE comparison between proposed pilot design and single pilot method in scenario 2}
	\label{fig:s2}
\end{figure}

We then test the radar sensing performances of FAOR, where only 1/8 symbols are used as pilot. Figure \ref{fig:bistaticee18} shows the estimation error of the radar sensing on the two targets, which gives both range error and speed error for the SNR from -30 dB to -10 dB. Range and speed estimation of target 2, which carries less power compared to target 1, remain stable only from around -20 dB. Less information is used in bistatic radar sensing leads to the performance degradation compared to monostatic radar.

In Figure \ref{fig:compareee}, we show the comparison of radar sensing performance between state-of-the-art radar sensing algorithm \cite{zhang2023radar} and FAOR. Results show that both algorithms performs well for target 1. While for target 2, FAOR outperforms the compared algorithm in both range and speed estimation at low SNR region.

\subsection{Evaluation on Channel Estimation}

\begin{table}[htb]
	\begin{center}
		\caption{Parameter setting}
		\addtolength{\tabcolsep}{-5pt}
		\scriptsize
		\begin{tabular}{|c|c|}
			\hline
			\textbf{Parameters}  &\textbf{Values} \\
			\hline
			Number of subcarriers N  &4096\\
			\hline
			Subcarrier interval &30 KHz\\
			\hline
			Sampling frequency  &122.88 MHz \\
			\hline
			Carrier frequency  &60 GHz  \\
			\hline
			Number of OTFS symbols used for sensing &100\\
			\hline
			Number of targets&3\\
			\hline
			Power levels of first reflected signals  &0 dB  \\
			\hline
			Power levels of second reflected signals&-3 dB\\
			\hline
			Power levels of third reflected signals&-6 dB\\
			\hline
			Speed for first target  & \makecell{Uniformly distributed in [5.4  37.8] km/hr}\\ 			
                \hline
			Speed for second target&\makecell{Uniformly distributed in [16.2  48.6] km/hr}\\
                \hline
			Speed for third target&\makecell{Uniformly distributed in [27  59.4] km/hr}\\			\hline
			Range of first target in S1 &Uniformly distributed in [3.66 18.3 ] m\\
			\hline
			Range of second target in S1 &Uniformly distributed in [8.54 23.2 ] m\\
			\hline
			Range of third target in S1 &Uniformly distributed in [13.4 28] m\\
	     	\hline
			Range of first target in S2 &Uniformly distributed in [3.66 125.73 ] m\\
			\hline
			Range of second target in S2 &Uniformly distributed in [8.54 130.6 ] m\\
			\hline
			Range of third target in S2 &Uniformly distributed in [13.4 135.5 ] m\\
			
			\hline
			
		\end{tabular}
		\label{tab:ce}
	\end{center}
\end{table}

The simulation parameters are listed in Table \ref{tab:ce}. We compare our proposed channel estimation algorithm with the single pilot channel estimation method described in \cite{raviteja2019embedded} under two scenarios: 
 \begin{itemize}
	\item Scenario 1 (S1): The delay spread of the targets remains within the allowed maximal delay. 
	\item Scenario 2 (S2): The delay spread of the targets sometimes exceeds the allowed maximal delay.
\end{itemize}

The single pilot aided channel estimation method proposed in \cite{raviteja2019embedded} adopts the following pilot design. Let $l_\tau$ and $k_v$ denote the taps corresponding to the maximum delay and Doppler value, respectively. In the delay Doppler domain, a single pilot symbol is placed in the center of a rectangular area of size $(2l_\tau+1)(2k_v+1)$, while the remaining symbols are guard symbols assigned with zeros. 

Our proposed pilot pattern is highly flexible, allowing for any shape and placement of pilot symbols within the delay Doppler domain. For comparison, we use the same rectangular pilot pattern as an example to evaluate our proposed channel estimation method with the single pilot channel estimation method.

In the simulation, the proposed pilot design utilizes a 55$\times$55 grid, with 3025 pilot symbols randomly assigned binary value of +1 or -1. For the single pilot channel estimation method, a single pilot symbol with a value of 55 is placed in the center of 55$\times$55 rectangular area, while all the remaining symbols are set to zeros. The overall power of the pilot symbols are the same in the two designs. Note that in the single pilot design, the PAPR is 44 dB, while in our proposed pilot design, the PAPR is only 13 dB. In both cases, the overhead of pilot is $\frac{55\times55}{4096\times100}=0.74\%<1\%$.

Fig \ref{fig:s1} shows the channel estimation mean square error (MSE) comparison between the methods in scenario 1. In the figure, $H$ denotes the combined channel estimation, the MSE of which can be written as,
\begin{equation}
	{\rm MSE}_{H}=\sum_{p=1}^{P}(h_p-\hat{h}_p)^2
\end{equation}
where $h_p$ is channel coefficient for target $p$, and $\hat{h}_p$ is the estimation of $h_p$.

From the \ref{fig:s1} we can see that the performance of our proposed method is close to that of single pilot design. However, our proposed method has much lower PAPR than the single pilot design, thus it is more suitable for actual implementation.

Fig \ref{fig:s2} shows the channel estimation MSE comparison between the methods in scenario 2. From Fig \ref{fig:s2} we can see that, when the delay spread of the targets sometimes exceeds the allowed maximal delay, our proposed method outperforms the single pilot method.

\section{CONCLUSION}

We have proposed the FAOR algorithm that can perform radar sensing in low complexity to detect the range and speed of the targets. It can be applied to not only monostatic radar but also bistatic radar with a much lower computational complexity compared to state-of-the-art radar sensing technology. We also have proposed new pilot designs for both sensing and communication, and channel estimation methods by using the detected time delays and Doppler frequencies of the targets after the radar sensing.

\bibliographystyle{IEEEtran}
\bibliography{FAORnew}


\end{document}